# Thin-film Al$_{0.30}$Ga$_{0.70}$As (111) as a 'flat' source of high-purity orthogonally polarized entangled photons

*Simon Stich[1], Vitaliy Sultanov[2,3], Trevor Blakie[4], Qingyu Shi[1],*

*Zbig Wasiliewski[4], Mikhail A. Belkin[1], and Maria Chekhova[2,3]*

[1]Walter Schottky Institute, Technical University of Munich, Garching, Germany

[2]Max-Planck Institute for the Science of Light, Erlangen, Germany

[3]Friedrich-Alexander Universität Erlangen-Nürnberg, Erlangen, Germany

[4]Waterloo Institute for Nanotechnology, University of Waterloo, Waterloo, Canada

ABSTRACT

Flat-optics platforms offer new opportunities for the generation of entangled photons by relaxing traditional phase-matching constraints, enabling the use of a broader range of nonlinear materials. Among these, gallium arsenide and aluminum gallium arsenide stand out for their exceptionally high second-order nonlinearities, but their conventional orientation (001) has limited their applicability for photon-pair generation. By transitioning to crystals with (111) surface orientation, we overcome these limitations. We demonstrate a flat-optics-based telecom-range SPDC source using Al$_{0.30}$Ga$_{0.70}$As that achieves a high photon-pair generation rate per pump power and bandwidth of up to 0.24 Hz/mW/nm. The choice of 30% aluminum concentration allowed us to reduce pump absorption and photoluminescence background for photon pairs generation at telecom wavelengths by at least an order of magnitude compared to that of GaAs. The specific layer orientation facilitates the generation of orthogonally polarized entangled photons, a prerequisite for polarization-entangled states. Rather than directly probing entanglement, we observe the effect of hidden polarization. Our results highlight AlGaAs (111) as a promising platform for scalable quantum photonic sources and shed light on nonclassical polarization effects accessible through flat-optics engineering.

## INTRODUCTION

Entangled photon sources are crucial for quantum optics, with applications ranging from quantum communication and metrology to foundational tests of quantum mechanics. Among the various methods to generate entangled photons, spontaneous parametric down-conversion (SPDC) in second-order nonlinear media remains one of the most widely used. Traditional SPDC sources rely on bulk nonlinear crystals such as lithium niobate (LN) or potassium titanyl phosphate (KTP), where efficient photon pair generation requires strict phase-matching conditions and long interaction lengths due to the relatively modest nonlinear coefficients ($\chi^{(2)} \sim 5 - 30 \frac{\text{pm}}{\text{V}}$) [1]. These constraints often limit design flexibility and the accessible range of entangled states. At the same time, the longer the crystal, the stricter the requirement of phase-matching, which ultimately narrows down the variety of two-photon states that can be generated. Through periodic poling of the crystal domain, quasi phase-matching is possible with these materials. Nevertheless, it remains technically challenging to obtain a proper periodic poling of crystals for any desired state. The direct generation of polarization-entangled photons with long crystals using birefringence or two consequent crystals with cross-oriented optical axes requires the compensation of the spatial or temporal walk-off to achieve a high purity of the generated state [2].

In contrast, flat-optics platforms — including thin films and metasurfaces — offer a new paradigm for nonlinear quantum photonics [3], [4]. Their small thickness naturally relaxes the phase-matching requirement thereby enabling simultaneous phase-matching across multiple wavelengths and nonlinear processes, making them inherently multifunctional—unlike conventional waveguides, which typically support only a limited set of phase-matched interactions. Furthermore, the relaxed phase-matching constraints allow the use of isotropic or otherwise phase-mismatch-limited materials that are typically unsuitable for conventional nonlinear optics. Photon pair generation using subwavelength-thin nonlinear materials including LN or zincblende structures such as gallium phosphide (GaP) [5], [6], as well as van der Waals [7], [8] and transition metal dichalcogenide (TMD) crystals [9], [10] has recently been demonstrated.

Crucially, the freedom to choose a wider range of materials opens the door to exploiting the exceptional nonlinear properties of gallium arsenide (GaAs), which possesses a second-order nonlinear susceptibility $\chi^{(2)} > 200 \frac{\text{pm}}{\text{V}}$ — the largest value found in bulk crystals and nearly an order of magnitude higher than typical bulk nonlinear crystals [11], [12].

One reason that prevented the use of GaAs for sub-wavelength SPDC sources is its predisposition to accidental single photon noise from spontaneous decay effects such as photoluminescence (PL), decreasing the purity of the generated photon pairs. This effect is especially dominant in thin films where the nonlinear interaction length $L$ is small. PL is an incoherent process; therefore, it scales linearly with $L$, while the pair-generation rate scales as $L^2$ and the former surpasses the latter when $L$ is too small [6], [13]. Fortunately, because PL photons only contribute to accidental coincidences rather than true coincidences, the real photon pair signal can still be observed when using low pump powers and extended acquisition times. In GaAs, PL is caused by pumping the material with photon energies larger than the bandgap. Note that in order to generate signal and idler photons at the telecom frequency band (1550nm, ~0.8eV) a pump wavelength of ~775nm (~1.6eV) is required, which is well above the bandgap energy of GaAs (1.42eV). By introducing aluminium, the bandgap energy can be increased above the pump photon energy, resulting in a significant reduction of PL. On the other hand, we must note that an increase of the aluminium fraction also comes with a decrease of $\chi^{(2)}$, as discussed in Ref. [12]. As a compromise, here we chose an aluminium fraction of 30%, increasing the bandgap energy to approximately 1.8 eV, while only moderately decreasing the nonlinear susceptibility by approximately 8%. The introduction of aluminium comes with the additional advantage of significantly reducing the optical losses at the pump wavelength, raising the optical damage threshold, thereby enabling higher pump powers which ultimately leads to an increased photon pair rate.

An additional challenge of using zincblende materials for nonlinear frequency mixing applications, even when phase-matching constraints have been addressed, is the peculiar nature of the $\chi^{(2)}$ tensor of crystals such as GaAs, AlGaAs and GaP grown on a regular (100) substrate. The only non-zero tensor elements for such materials are given by the crystal point group $\bar{4}3m$:

$$\chi^{(2)}_{xyz} = \chi^{(2)}_{yxz} = \chi^{(2)}_{xzy} = \chi^{(2)}_{yzx} = \chi^{(2)}_{zyx} = \chi^{(2)}_{zxy}.$$

From this, it becomes apparent that in order to promote any second-order nonlinear process such as SPDC or second-harmonic generation (SHG), the involved electric fields must possess all three spatial components, which cannot be achieved by pumping along a crystallographic axis (the growth direction).

Several strategies have been developed to exploit the strong nonlinearity of zincblende materials under normally incident pump light, despite their unfavorable crystal symmetry in standard orientations. One such approach involves growing the material along crystallographic directions other than the conventional [100]. For instance, the GaP films reported in Refs. [5], [6] were fabricated with their surface normal tilted 15° away from the [100] direction, ensuring that the pump electric field has projections on all principal crystallographic axes. Alternatively, Ref. [14] demonstrates the use of GaAs nanowires grown along the [111] direction, which are then laid flat on a substrate to align the nonlinear tensor components with the optical field and effectively utilize the material's high nonlinearity.

Nanostructuring a thin film into a metasurface offers an additional strategy to tackle this challenge. Recently, GaAs and InGaP thin films grown on standard (100) or (110) substrates have been patterned into metasurfaces to enable efficient nonlinear interactions [15], [16], [17]. Metasurface design circumvents the orientation limitation by introducing nanoresonators that enable the electric field to interact with the $\chi^{(2)}$ tensor through optical resonances, effectively unlocking SPDC emission at normal incidence.

We present an alternative approach to utilize the strong nonlinear properties found in III-V semiconductors. Instead of fabricating the thin films from regular (100) substrates, substrates with different orientations, such as (111), provide the possibility to directly obtain polarization-entangled photon pairs from a thin nonlinear film with the pump beam impinging along the growth direction. By using (111) substrates, the second-order nonlinear susceptibility tensor $\chi^{(2)}$ will change with respect to the laboratory reference coordinate system. One possible rotation, where the laboratory $z$-axis is aligned with the crystallographic (111) direction and the laboratory $x'$- and $y'$-axis are oriented along the crystals ($2\bar{1}\bar{1}$) and ($01\bar{1}$) direction, reveals the following in-plane components of the rotated nonlinear susceptibility tensor $\chi^{(2)\prime}$:

$$-\chi^{(2)}_{x'x'x'} = \chi^{(2)}_{x'y'y'} = \chi^{(2)}_{y'y'x'} = \chi^{(2)}_{y'x'y'} = \sqrt{\frac{2}{3}}\chi^{(2)}_{xyz}, \qquad (1)$$

enabling nonlinear interaction with electric fields only polarized in plane (i.e., $x'$ or $y'$). For the sake of brevity, we have omitted the out-of-plane components of the tensor.

It is important to emphasize that, unlike x-cut LN presented in Ref. [6], where the largest nonlinear tensor component $d_{33}$ is exploited to generate co-polarized photon pairs, (111)-oriented zincblende layers inherently produce pairs of orthogonally polarized photons, making them directly suitable for the generation of polarization-entangled states. Moreover, GaAs substrates with (111) orientation are readily available and reliable epitaxial growth of high quality AlGaAs films with various aluminium fractions can be achieved [18].

In summary, by combining the high nonlinearity of AlGaAs, a tailored (111) crystal orientation, and the relaxed phase-matching conditions of flat optics, we realize an efficient and versatile platform for nonlinear quantum optics. The result is a compact source that generates photon pairs with a high rate, low noise background, and well-defined polarization structure.

METHODS

Our sample consists of an $Al_{0.30}Ga_{0.70}As$ (111) film of thickness 1.8 μm adhesively bonded to a double-side-polished c-cut sapphire host substrate using Benzocyclobutene (BCB). For comparison, a similar reference sample made of a 250-nm-thick film of GaAs (111) adhesively bonded to a c-cut sapphire substrate with BCB was fabricated. Both semiconductor films were initially grown on GaAs substrates with (111) orientation. The epilayers include the semiconductor films ($Al_{0.30}Ga_{0.70}As$ or GaAs) grown on top of an etch stop layer made of $Al_{0.55}Ga_{0.45}As$. The wafer was adhesively bonded epi side-down to a c-cut sapphire host substrate using BCB. The native GaAs substrate was then removed by first lapping most of it and then selective chemical wet etching it with a mixture of citric acid and $H_2O_2$ down to the $Al_{0.55}Ga_{0.45}As$ etch-stop layer [19]. Finally, the $Al_{0.55}Ga_{0.45}As$ etch stop layer was selectively removed with a concentrated hydrofluoric acid.

Both samples were characterized in the same setup, which is depicted in Fig. 1(a). We pumped SPDC at a wavelength of 788 nm with a continuous-wave pigtail diode laser. As the laser polarization can be subject to polarization scrambling in the fiber, a Glan prism was used to ensure linear polarization (not shown in Fig. 1(a)). Before focusing, a half-wave plate (HWP) was inserted to rotate the pump polarization. The sample was pumped through its sapphire substrate and the emitted SPDC radiation was collected using a lens with 0.7 numerical aperture (NA). To block any unwanted pump radiation, a long-pass (LP) filter with a cutoff at 1400 nm was placed after the collecting lens. To characterize the photon pair polarization, we additionally inserted a series of a HWP and a polarizing beamsplitter (PBS) in the form of a Glan-laser prism into the beam path. Using another lens, the photon pairs were focused into the fiber and then sent through a 50/50 fiber beamsplitter to two superconducting nanowire single-photon detectors (SNSPDs). Here, the coincidences were counted using a time tagger device. To characterize the emitted SPDC spectrum, a dispersive fiber spool was attached for this specific measurement between one output of the 50/50 beamsplitter and one SNSPD to perform time-of-flight spectroscopy of photon pairs [20].

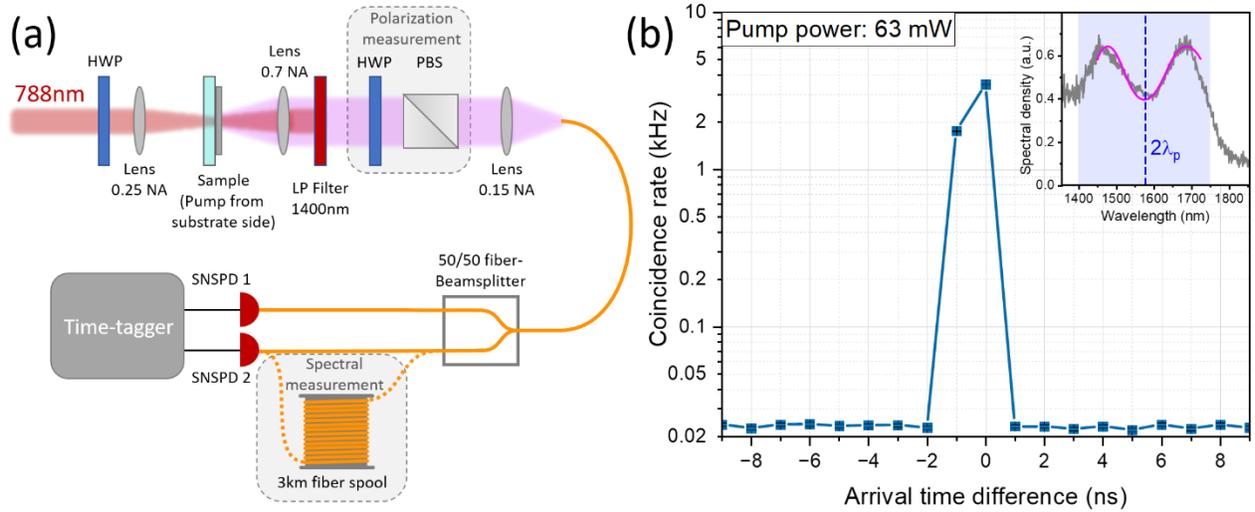

Figure 1. (a) Measurement setup for characterizing the emitted photon pairs. Polarization analysis was performed by adding an extra half-wave plate (HWP) and polarizing beam splitter (PBS) in the beam path. To characterize the spectrum of the photon pairs, one output arm of the 50/50 fiber beamsplitter was connected to a 3 km long dispersive fiber. Measurements of the pure photon pair rate were conducted without these additional components. (b) Coincidences obtained from the 1.8 μm thick $Al_{0.30}Ga_{0.70}As$ film at a pump power of 63 mW with a maximum photon pair rate of 5.2 kHz (corrected for accidentals). The inset shows the measured (grey) and theoretical (magenta) spectrum of the emitted photon pairs. The oscillations are due to the Fabry-Pérot fringes of the thin film. The blue dashed line denotes the degenerate wavelength (twice the pump wavelength $\lambda_p$) of the photon pairs. The shaded blue area represents the detection bandwidth (350nm) of the measurement setup, bounded on the short-wavelength side by a long-pass filter with a cutoff at 1400 nm, and on the long-wavelength side by the SNSPDs' detection limit at 1750 nm.

RESULTS AND DISCUSSION

In the initial experiment, the maximum achievable photon pair rate of the AlGaAs thin film was evaluated. For this measurement, both the HWP and the PBS were omitted, allowing detection of photons with all polarizations by the SNSPDs. The sample was pumped using the highest available power of 63mW, with the pump polarization adjusted to maximize the photon pair rate. The recorded coincidence counts are presented in Fig. 1(b), the peak showing a photon pair rate of 5.2 kHz under these conditions. The corresponding photon pair spectrum is displayed in grey in the inset of Fig. 1(b). As anticipated—unlike in bulk crystals—the emitted photon pairs span a broad spectral range centered around the degenerate SPDC wavelength, $2\lambda_p$ (indicated by the dashed blue vertical line). This measured spectrum is constrained by the 350 nm detection bandwidth of the setup (shaded blue area), which is limited on the short-wavelength side by a long-pass filter with a 1400 nm cutoff (used to block the pump) and on the long-wavelength side by the SNSPDs' detection limit at 1750 nm. The observed oscillations in the spectrum are Fabry-Pérot fringes [21], resulting from reflections at the layer interfaces, and show excellent agreement with theoretical predictions (magenta). Based on these values, we obtain a record-high photon pair generation rate, normalized to pump power and bandwidth, of 0.24 Hz/mW/nm, exceeding previously reported values for thin-film sources. For comparison, the highest rate previously reported for a LN thin film was 0.006 Hz/mW/nm [6], while even highly resonant (110) GaAs metasurfaces reached only 0.025 Hz/mW/nm [16]. It should be noted, however, that the thickness of our AlGaAs film is considerably greater than that of the aforementioned sources, which contributes to the high photon pair rate. When normalized to the square of the nonlinear material's thickness—a metric that more fairly compares nonlinear interaction efficiencies—our AlGaAs film performs only slightly below the GaAs metasurface reported in Ref. [16], which, to the best of our knowledge, currently is the brightest thin SPDC source.

To directly compare AlGaAs with GaAs, similar measurements were performed on both thin films under varying pump powers. The pump power was measured immediately before the sample using a photodiode power sensor. Coincidence peaks for both films, pumped at 9mW, are displayed in Fig. 2(a). As expected, the thicker AlGaAs film exhibits a significantly higher photon pair rate, reflected in the taller peak at 0 ns. Despite its greater thickness, the AlGaAs film shows a lower background noise level, suggesting a reduction in photoluminescence (PL)-induced spontaneous emission.

To further investigate this background contribution from PL, Fig. 2(b) presents the detected photon rates for one of the SNSPDs across varying pump powers, along with corresponding linear fits. Because the rate of the PL, which is an incoherent effect, scales linearly with the thickness of the layer, and PL constitutes most of the single-photon signal, the plotted rate is normalized to the sample thickness. As anticipated, the GaAs film yields substantially higher photon rates due to its stronger PL level compared to AlGaAs.

Beyond the lower noise, the introduction of aluminum also significantly improves the damage threshold of the material. The GaAs sample exhibited visible degradation at pump powers below 10 mW, whereas the AlGaAs film withstood the full range of pump powers without damage. This difference is also reflected in the photon pair rate measurements shown in Fig. 2(c). Due to differences in film thickness, the data were normalized accordingly. When the emitted spectrum exceeds the detection bandwidth (demonstrated in the inset of Fig. 1(b)), the observed photon pair rate scales as $L^2 \times \text{sinc}^2(\Delta k L/2)$, where $\Delta k$ denotes the phase mismatch for the considered SPDC process [22], i.e., for the degenerate case $\Delta k \approx k_\text{p} - 2k_\text{s} \approx (1.8\mu\text{m})^{-1}$. While phase mismatch is negligible for very thin films such as GaAs, it becomes relevant for the Al$_{0.30}$Ga$_{0.70}$As sample, whose thickness is on the order of $1/\Delta k$. The photon pair rates from AlGaAs show a linear dependence on pump power, as evidenced by their strong agreement with the linear fit. In contrast, the GaAs data deviate from linearity even at relatively low pump powers due to the aforementioned sample degradation as depicted in the inset of Fig. 2(c).

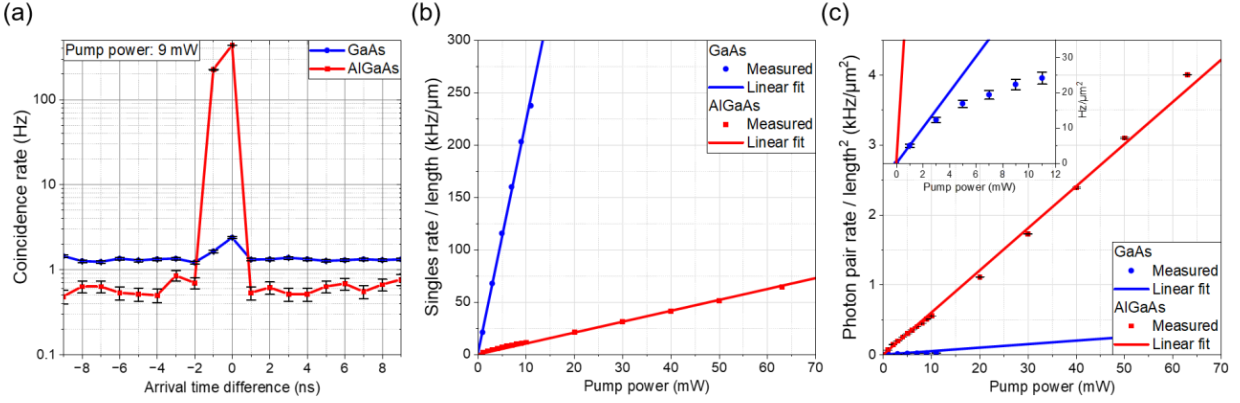

Figure 2. (a) Coincidence plot for both samples at 9 mW pump power. (b) Singles rates obtained for both samples from one of the two SNSPDs for varying pump powers along with linear fits. The rates are normalized to the samples thicknesses L. (c) Photon pair rates from both samples for varying pump powers with corresponding linear fits. The photon pair rates are normalized by a factor $L^2 \times \text{sinc}^2(\Delta k L/2)$ to account for the different thicknesses and the phase mismatch. The inset highlights the deviation from the linear trend for the GaAs sample at high pump powers.

The differing slopes of the two linear fits in Fig. 2(c) are surprising, since the nonlinear susceptibility of AlGaAs is expected to diminish slightly at moderate aluminum concentrations (see Ref. [12]). We believe this discrepancy arises primarily from optically induced damage even at the lowest considered pump powers and the challenges of efficiently coupling the photon pairs from the GaAs film into the fiber—its low photon pair output and elevated background noise made optimal alignment far more difficult.

From Fig. 2 we can estimate the ratio between SPDC and PL photon emission rates, which determines the purity of the two-photon state [13]. From panel (a), we infer the rate of coincidences per pump power as 67 Hz/mW. With the detection efficiency by a single detector being 10-20% (which includes 50% loss at the beam splitter, other optical losses, and the coupling into the single-mode fibre) [13], the rate of SPDC photons detection by the second detector can be estimated as 0.5 ± 0.2 kHz/mW. According to panel (b), the total singles rate per pump power is 1.8k Hz/mW. This rough estimate shows a remarkably high, for thin sources, fraction of SPDC photons: 30 ± 10 %. For comparison, in a similar setup with a 7μm lithium niobate (LN) film, only 1% of detected photons were from SPDC [13].

Our experiments show that the AlGaAs film performs almost as well as the highly resonant metasurfaces made from GaAs with record brightness demonstrated in Ref. [16] while, at the same time, keeping experimental noise at a minimum. For comparison, Table 1 shows the photon pair rates and the fractions of SPDC photons for recently demonstrated thin SPDC sources targeting the telecom wavelength band. While metasurfaces, generally, show higher photon pair rates per spectral bandwidth than unstructured films, their PL level is extremely high, exceeding the level of SPDC photons by three to four orders of magnitude. Meanwhile, our AlGaAs film features almost the same brightness as the 'most bright' metasurface and, at the same time, the highest SPDC-to-PL ratio, far exceeding the values obtained for any metasurfaces and films.

| Source | Thickness [μm] | Wavelength [nm] | Photon pair detection rate per power and bandwidth [$s^{-1}$ m$W^{-1}$ n$m^{-1}$] | Photon pair detection rate per power, bandwidth, thickness$^2$ [$s^{-1}$ m$W^{-1}$ n$m^{-1}$ μ$m^{-2}$] | Fraction of SPDC photons | Reference |
|---|---|---|---|---|---|---|
| GaP layer | 0.4 | 1275 | $3 \cdot 10^{-4}$ | $1.9 \cdot 10^{-3}$ | 0.3% | [5] |
| LN layer | 0.3 | 1370 | $6 \cdot 10^{-3}$ | $6.7 \cdot 10^{-2}$ | ~1% | [6] |
| LN Mie metasurface | 0.7 | 1570 | $8 \cdot 10^{-3}$ | $1.6 \cdot 10^{-2}$ | 0.3% | [23] |
| GaAs BIC (001) metasurface A | 0.5 | 1450 | $3 \cdot 10^{-3}$ | $1.2 \cdot 10^{-2}$ | 0.01% | [15] and SI |
| InGaP (110) metasurface | 0.5 | 1560 | $2 \cdot 10^{-3}$ | $8 \cdot 10^{-3}$ | unknown | [17] |
| GaAs MD-qBIC (110) metasurface B | 0.5 | 1580 | $2.5 \cdot 10^{-2}$ | $1 \cdot 10^{-1}$ | unknown | [16] |

| AlGaAs layer | 1.8 | 1576 | $2.4 \cdot 10^{-1}$ | $8 \cdot 10^{-2}$ | 30% | This work |

Table 1: Comparison of different thin SPDC sources demonstrated recently. The green shading denotes thin films while the yellow shading indicates metasurfaces composed of nanoresonators. For GaP and LN layers and the LN Mie metasurface, the singles rate was inferred from the rate of accidental coincidences. The rate of singles caused by SPDC was estimated from the rate of coincidences assuming a 10% detection efficiency.

We further investigated polarization correlations in photon pairs produced via SPDC by inserting a single HWP and polarizer in the optical path prior to beam splitting into the two detection arms. This setup allowed us to rotate the polarization state of the photon pairs and project it onto different axes before spatial separation and detection. Coincidence counts were measured as a function of the photon polarization angle $\theta$ (twice the HWP angle), which effectively rotates the polarization basis for both photons simultaneously.

The measured coincidence rate acquired at 50 mW pump power is plotted in red in Fig. 3 and shows a clear sinusoidal dependence on the photon polarization angle, following a $\sin^2(\theta)$ pattern (shown in orange dashed), consistent with the expected behavior for orthogonally polarized photon pairs. Furthermore, the rate of accidental coincidences (shown in blue)—arising from uncorrelated photons—remained flat, confirming that the observed modulation in true coincidences stems from genuine polarization correlations. Note that with SPDC photons constituting a considerable fraction of all single counts, the absence of modulation can be attributed to SPDC.
This disparity between the polarization-insensitive accidental rates and the strongly polarization-dependent coincidence signal illustrates the phenomenon of hidden polarization [24], [25], [26]. While individual photons appear unpolarized when measured alone, their joint statistics reveal strong polarization correlations. These correlations are not captured by first-order polarization observables (such as the Stokes parameters of each beam) but emerge in second-order (coincidence) measurements, characteristic of non-classically correlated states.

A state of orthogonally polarized photon pairs,

$$|\Psi\rangle = |1\rangle_H |1\rangle_V, \qquad (2)$$

where $|1\rangle_{H,V}$ is a single-photon state in horizontal or vertical polarization mode, can be converted into a polarization-entangled state by sending it into a non-polarizing beam splitter and discarding events where both photons of a pair exit from the same output [5]. At higher pump power, with increased flux of photon pairs, the state features polarization squeezing, i.e., reduced noise in one of the Stokes observables [24], [28]. Due to the polarization noise reduction, polarization-squeezed light provides enhanced sensitivity in polarization measurements [27], [29].

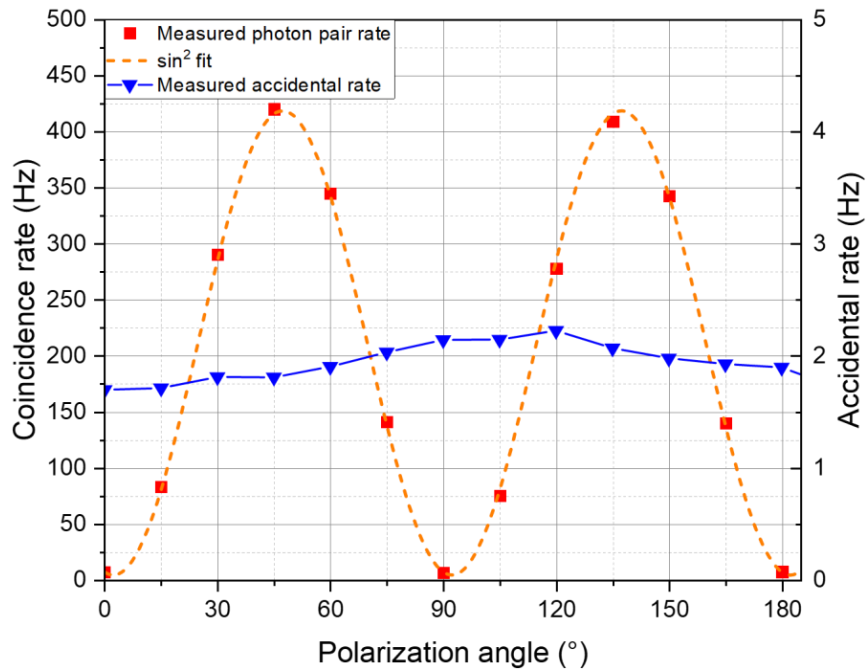

Figure 3: Measured photon pair coincidence rate for each polarization angle. The four-fold symmetry indicates hidden polarization and can be used to form a maximally entangled Bell state.

Therefore, AlGaAs offers not only one the strongest second-order nonlinear susceptibilities among all crystals used for SPDC but also allows for the generation of polarization-entangled photons. Notably, SPDC in an AlGaAs film with (111) orientation will produce state (2) under any pump wavelength, due to the automatically satisfied phase matching, and under any pump polarization, due to the structure of the nonlinear tensor. Together with the high purity of the two-photon state, this makes thin AlGaAs a promising platform for quantum light engineering.

CONCLUSION

In conclusion, we have demonstrated that (111)-oriented GaAs/AlGaAs thin films allow effective use of the material's large second-order nonlinearity for SPDC under normal-incidence pumping. This enables the realization of compact photon pair sources with significantly higher generation rates than those achieved using conventional materials of similar thickness, such as LN or GaP. The introduction of a moderate aluminum content does not notably deteriorate the nonlinear susceptibility, thereby preserving the strong nonlinear properties of the AlGaAs/GaAs material system. Our AlGaAs film achieves a normalized photon pair rate of 0.24 Hz/mW/nm, exceeding previously reported values for thin-film sources [6] including metasurfaces [16]. These results highlight the potential of (111)-grown AlGaAs as a leading platform for efficient, broadband, and compact photon pair sources.

In addition, AlGaAs with moderate aluminum content proves to be advantageous over GaAs for the application of entangled photon sources in the telecom band due to the reduced optical losses at the pump wavelength, enabling higher pump powers and thereby higher photon pair rates. The lower losses additionally reduce the generated photoluminescence, which usually strongly contaminates two-photon light produced via SPDC in thin films and metasurfaces. While the fraction of SPDC photons in such experiments typically does not exceed 1%, here about 30% of all photons are photon pairs. Thin-film AlGaAs is thus a source of high-purity entangled photons.

Finally, due to the symmetry of AlGaAs tensor and due to the automatic satisfaction of phase matching, SPDC in a (111) film generates pairs of orthogonally polarized photons, manifesting the 'hidden polarization' effect and at stronger pumping, potentially also polarization squeezing. The relaxed phase-matching conditions furthermore permit free choice of pump wavelength if required, allowing for even broader spectral coverage.

AUTHOR INFORMATION

**Corresponding Author**

Maria Chekhova, email: maria.chekhova@mpl.mpg.de

Simon Stich, email: simon.stich@wsi.tum.de

Mikhail A. Belkin, email: mikhail.belkin@wsi.tum.de

**Author Contributions**

The manuscript was written through contributions of all authors. All authors have given approval to the final version of the manuscript.

**Funding Sources**

This work was funded by Deutsche Forschungsgemeinschaft (DFG, German Research Foundation), Project 311185701, and by Bundesministerium fuer Bildung und Forschung (BMBF, German Ministry of Education and Research), Project 13N16933.